\documentclass[10pt]{article}
\usepackage{feynmf}

\begin{document}

\title{Two-Pion Decay Widths of Excited Charm Mesons}

\author{T.A. L\"ahde$\,$\footnote{talahde@pcu.helsinki.fi}$\:$
and$\:\:$D.O. Riska$\,$\footnote{riska@pcu.helsinki.fi} 
\\ \vspace{0.3cm} \\
{\normalsize \it Helsinki Institute of Physics
and Department of Physics}\\
{\normalsize\it POB 9, 00014 University of Helsinki, Finland} }

\date{}
\maketitle
\thispagestyle{empty}

\begin{abstract}
The widths for $\pi\pi$ decay of the $L=1$ charm mesons are calculated
by describing the pion coupling to light constituents quarks by the 
lowest order chiral interaction. The wavefunctions of the charm
mesons are obtained as solutions to the covariant Blankenbecler-Sugar
equation. These solutions correspond to an interaction Hamiltonian modeled 
as the sum of a linear scalar confining and a screened one-gluon exchange 
(OGE) interaction. This interaction induces a two-quark contribution to 
the amplitude for two-pion decay, which is found to interfere 
destructively with the single quark amplitude. For the currently known 
$L=1$ $D$ mesons, the total $\pi\pi$ decay widths are found to \mbox{be 
$\sim 1$ MeV} for the $D_1(2420)$ and $\sim 3$ MeV for the $D^*_2(2460)$ 
if the axial coupling of the constituent quark is taken to be $g_A^q=1$. 
The as yet undiscovered spin singlet $D_1^*$ state is predicted to have a 
larger width of \mbox{7 - 10 MeV} for $\pi\pi$ decay. 
\end{abstract}
\newpage

\section{Introduction}

The pion decay widths of excited charm mesons, the $D$ mesons, which are 
formed of one light flavor quark (antiquark) and a charm antiquark (quark) 
provide the best   
observables for determining the coupling of pions to light constituent
quarks, as the pions do not couple to the charm quarks. The coupling
strength of the pion to the light flavor constituent quark is determined
by the axial coupling constant $g_A^q$ of the quark. The theoretically 
indicated values for $g_A^q$~\cite{Wein1,Wein2,Dicus,Dannbom}, which fall 
in the range 0.75 to 1.0, have been shown to be consistent with extant 
empirical information on the single pion decay widths of the ground state 
and excited charm mesons~\cite{Goity,TL3}. The uncertainty margin derives 
from the fact that only upper limits have been set 
experimentally~\cite{PDG} on the widths of the $D^*$ mesons, while the
total decay widths of the orbitally excited $L=1$ $D_1(2420)$ and 
$D_2^*(2460)$ charm meson states are known only within a very wide 
uncertainty range. As the latter lie well above
the threshold, not only for single pion but also for two pion decay, it
would be particularly instructive to obtain theoretical 
predictions for as well as empirical information on the branching ratios 
for the latter decay modes.

We here report a calculation of the two-pion decay widths of
the excited $L=1$ charm meson states, by extending a similar earlier
calculation of the widths of their single pion decay modes~\cite{TL3}.
The charm mesons consisting of a light quark (or antiquark) $q$ and a 
heavy antiquark (or quark) $\bar Q$ are here treated as relativistic 
two-particle systems described by wavefunctions which have been obtained 
as solutions to the covariant Blankenbecler-Sugar 
equation~\cite{Blsu,Tavkh} in ref.~\cite{TL2}. The interaction between 
the $q$ and the $\bar Q$
(or the $\bar q$ and the $ Q$) is described as a combined linear
confining and screened one-gluon exchange interaction, with parameters
determined so as to obtain a satisfactory prediction of the empirically 
known part of the $D$ meson spectrum~\cite{TL2}.

The model for emission of two pions from the light constituent quarks
employed is the conventional chiral pion-quark pseudovector coupling 
model, which includes a Weinberg-Tomozawa type term for constituent 
quarks. The only parameter in this amplitude is the axial coupling of the 
light constituent quarks $g_A^q$, as the quark mass is fixed by
the Hamiltonian model used to determine the $D$ meson spectrum and 
wavefunctions. In addition to the single quark amplitude for two-pion 
emission, we also consider the interaction current contribution to the 
vector current of the $q\bar Q$ system, which is associated with 
intermediate antiquark terms. This exchange current contribution to the 
Weinberg-Tomozawa interaction is found to interfere destructively 
with the single quark amplitude and to bring about a reduction of the
calculated width for the two pion decay of the $L=1$ $D$ mesons by about 
25\%. 

The amplitude for two-pion decay is found to be mainly spin-independent, 
which is in accordance with the current empirical information on the 
$\pi\pi$ decays of the analogous strange $K$ mesons. The $\pi\pi$ decay 
width of the $D_2^*$ meson is predicted to be about 3 MeV. For the $D_1$ 
meson, the corresponding number is found to be 1 MeV. Estimates are also 
given for the $\pi\pi$ decay widths of
the hitherto undiscovered $D$ meson states with $L=1$, i.e. the spin 
triplet $D_0^*$ and the spin singlet $D_1^*$. The masses of these two 
states are predicted in ref.~\cite{TL2} to be $2340$ and $2390$ MeV 
respectively. The $\pi\pi$ decay width of the $D_0^*$ is here found to be 
small (of the order one tenth of an MeV, whereas that of the $D_1^*$ is 
predicted to be about \mbox{7 MeV.} This variation is a natural 
consequence of 
the rather small phase space available for these decays, combined with the 
significant spin-spin and spin-orbit splittings between various $D$ meson 
states.

This paper is divided into 4 sections. In section 2 the operator for
two-pion emission from a single quark is described along with a
derivation of the corresponding decay widths. In section 3 the
exchange current contribution to the two-pion decay width is calculated. 
Section 4 contains a concluding discussion.

\newpage

\section{The Decay Rate for $\pi\pi$ decay}

\subsection{Single Quark Amplitude for Two-Pion Decay}

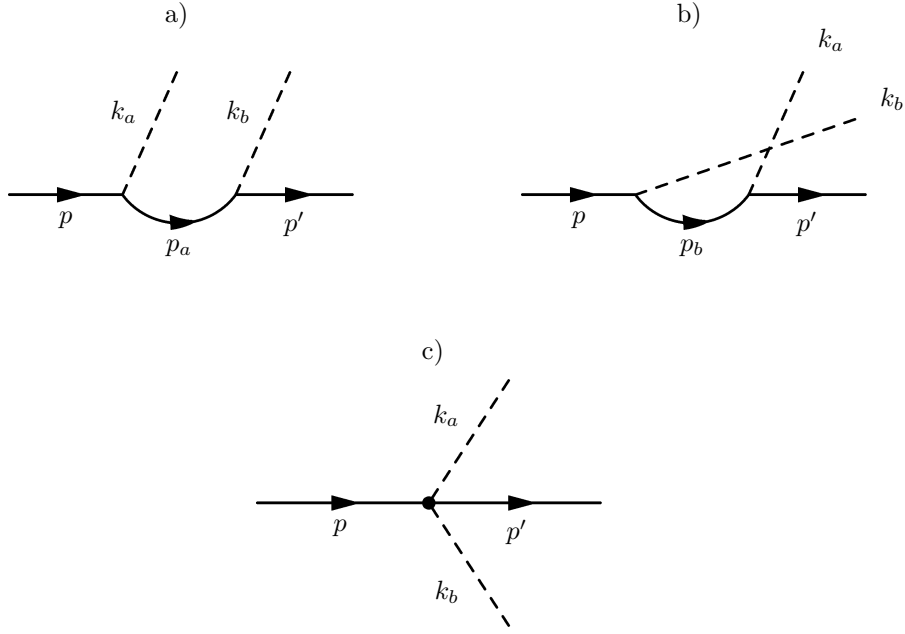
\begin{figure}[h!]
\begin{center}
\begin{tabular}{c c c} \\
a) && b) \\ \\ 
\begin{fmffile}{Born}   
\begin{fmfgraph*}(130,100) \fmfpen{thin}
\fmfleft{i1}
\fmftop{o1,o2}
\fmfright{o3}
\fmf{fermion,label=$p$}{i1,v1}
\fmf{fermion,label=$p_a$,right=.5,tension=.5}{v1,v2}
\fmf{fermion,label=$p'$}{v2,o3}
\fmf{dashes,label=$k_a$}{v1,o1}
\fmf{dashes,label=$k_b$}{v2,o2}
\fmfforce{(0,.5h)}{i1}
\fmfforce{(0.33w,.5h)}{v1}
\fmfforce{(0.66w,.5h)}{v2}
\fmfforce{(w,.5h)}{o3}
\fmfforce{(0.5w,h)}{o1}
\fmfforce{(0.83w,h)}{o2}
\end{fmfgraph*}
\end{fmffile}
& \quad \quad \quad \quad & 
\begin{fmffile}{CBorn}
\begin{fmfgraph*}(130,100) \fmfpen{thin}
\fmfleft{i1}
\fmftop{o1,o2}
\fmfright{o3}
\fmf{fermion,label=$p$}{i1,v1}
\fmf{fermion,label=$p_b$,right=.5,tension=.5}{v1,v2}
\fmf{fermion,label=$p'$}{v2,o3}
\fmf{dashes}{v2,o1}
\fmf{dashes}{v1,o2}
\fmfforce{(0,.5h)}{i1}
\fmfforce{(0.33w,.5h)}{v1}
\fmfforce{(0.66w,.5h)}{v2}
\fmfforce{(w,.5h)}{o3}
\fmfforce{(0.83w,h)}{o1}
\fmfforce{(w,.8h)}{o2}
\fmflabel{$k_a$}{o1}
\fmflabel{$k_b$}{o2}
\end{fmfgraph*}
\end{fmffile} \\
& c) & \\
\end{tabular}
\begin{fmffile}{WT}   
\begin{fmfgraph*}(130,100) \fmfpen{thin}
\fmfleft{i1}
\fmftop{o1}
\fmfbottom{o2}
\fmfright{o3}
\fmf{fermion,label=$p$}{i1,v1}
\fmf{fermion,label=$p'$}{v1,o3}
\fmf{dashes,label=$k_a$}{v1,o1}
\fmf{dashes,label=$k_b$}{v1,o2}
\fmfdot{v1}
\fmfforce{(0,.5h)}{i1}
\fmfforce{(.5w,.5h)}{v1}
\fmfforce{(w,.5h)}{o3}
\fmfforce{(.75w,h)}{o1}
\fmfforce{(.75w,0)}{o2}
\end{fmfgraph*}
\end{fmffile}
\caption{Feynman diagrams that correspond to the interaction Lagrangians of 
eqs.~(\ref{lagr}) and~(\ref{Tomoz}). The diagrams a) and b) describe the 
Born and crossed Born amplitudes and the diagram c) 
describes the contact vertex that is associated with the Weinberg-Tomozawa 
Lagrangian, eq.~(\ref{Tomoz}). In these diagrams $p_a = p-k_a$, $p_b = 
p-k_b$ and $k_a + k_b = p-p'$.}
\label{diagram}
\end{center}
\end{figure}

The emission of two pions from a $D$ meson may be described by the 
pseudovector Lagrangian, which constitutes the lowest order chiral 
coupling for pions to constituent quarks:

\begin{equation}
{\cal L}=i{g_A^q\over 2f_\pi}\bar\psi_q\,\gamma_5\gamma_\mu
\,\partial_\mu\,\vec\phi_\pi\cdot\vec \tau\,\psi_q.
\label{lagr}
\end{equation}
Here $g_A^q$ denotes the 
axial coupling constant of pions to light constituent quarks, and $f_\pi$ 
is the pion decay constant, the empirical value of which is 93 MeV. The 
axial coupling constant is conventionally taken to be equal to, 
or somewhat less than, unity~\cite{Wein1,Wein2}. This coupling gives
rise to Born term and crossed Born amplitudes 
of conventional form, Fig.~\ref{diagram}~(a,b), for the emission of two 
pions from an interacting constituent quark. 

The chiral model for the 
two-pion decay amplitude is completed by the 
Weinberg-Tomozawa~(WT) interaction, which is described by the Lagrangian

\begin{equation}
{\cal L}_{\mathrm{WT}} =
-\frac{i}{4f_{\pi}^2}\bar\psi_q\,\gamma_{\mu}\,
\vec\tau\cdot\vec\phi_{\pi}\times\partial_{\mu}
\vec\phi_{\pi}\,\psi_q.
\label{Tomoz}
\end{equation}
This interaction leads to the contact amplitude that is described
by the diagram~c) in Fig.~\ref{diagram}. The Born amplitude 
(Fig.~\ref{diagram}a) may be expressed as

\begin{equation}
T_{\mathrm B} =
-i\left(\frac{g_A^q}{2f_{\pi}}\right)^2 \bar\psi_q \:\gamma_5\gamma\cdot 
k_b\: \frac{1}{\gamma\cdot p_a -im_q}\:\gamma_5\gamma\cdot 
k_a\psi_q\:\tau_b\tau_a.
\label{Born1}
\end{equation}  
Here $m_q$ denotes the mass of the light constituent quark, which in 
ref.~\cite{TL2} was obtained as 450 MeV, as this value led to
an optimal description of the charm meson spectrum within the model 
considered. Similarly the crossed Born amplitude (Fig.~\ref{diagram}b) 
takes the form

\begin{equation}
T_{\mathrm{CB}} =
-i\left(\frac{g_A^q}{2f_{\pi}}\right)^2 \bar\psi_q\:\gamma_5\gamma\cdot 
k_a\: \frac{1}{\gamma\cdot p_b -im_q}\:\gamma_5\gamma\cdot 
k_b\:\psi_q\tau_a\tau_b.
\label{Born2}
\end{equation}  
The general isospin decomposition of the two-pion emission
amplitude for constituent quarks is, in analogy with that
for nucleons,

\begin{equation}
T = \delta_{ab}T^+ + \frac{1}{2}[\tau_b,\tau_a] T^-.
\label{plmi}
\end{equation}
The general invariant expression 
for for the amplitudes $T^+$ and $T^-$ is in turn 

\begin{equation}
T^\pm = \bar u(p') \left(A^\pm - i\gamma\cdot Q B^\pm\right) u(p).
\label{ampl}
\end{equation}
In these expressions, the four-vector $Q$ denotes the combination $Q = 
(k_b - k_a)/2$, where $k_a$ and $k_b$ are the four-momenta of the emitted 
pions. The commutator of the SU(2) generators may be expressed as 
$[\tau_b,\tau_a] = 2i\epsilon_{abc}\tau_c$. In this notation the Born, 
crossed Born and Weinberg-Tomozawa amplitudes are, respectively

\begin{equation}
T_{\mathrm B} = -i\left(\frac{g_A^q}{2f_{\pi}}\right)^2
\left(-\gamma\cdot Q + 2im_q + 4m_q^2\frac{\gamma\cdot Q}{p_a^2 + 
m_q^2}\right)
\left(\delta_{ba} + \frac{1}{2}\left[\tau_b,\tau_a\right]\right),
\end{equation}

\begin{equation}
T_{\mathrm {CB}} = -i\left(\frac{g_A^q}{2f_{\pi}}\right)^2
\left(\gamma\cdot Q + 2im_q - 4m_q^2\frac{\gamma\cdot Q}{p_b^2 + 
m_q^2}\right)
\left(\delta_{ba} - \frac{1}{2}\left[\tau_b,\tau_a\right]\right),
\end{equation}

\begin{equation}
T_{\mathrm{WT}} = - i\gamma\cdot Q \frac{1}{2f_{\pi}^2}
\frac{1}{2}[\tau_b,\tau_a].
\end{equation}

Comparison of these amplitudes 
with eq.~(\ref{ampl}) yields the desired expressions for the 
sub-amplitudes $A^\pm$ and $B^\pm$, which are

\begin{eqnarray}
A^+ &=& \left(\frac{g_A^q}{2f_{\pi}}\right)^2 4m_q, \\
A^- &=& 0, \\
B^+ &=& -\left(\frac{g_A^q}{2f_{\pi}}\right)^2 4m_q^2
\left[\frac{1}{s - m_q^2} - \frac{1}{u - m_q^2}\right], \\
B^- &=& -\left(\frac{g_A^q}{2f_{\pi}}\right)^2
\left(2 + 4m_q^2\left[\frac{1}{s - m_q^2} + \frac{1}{u - m_q^2}\right]
\right) + \frac{1}{2f_\pi^2}. \label{B-}
\end{eqnarray}
Here the identities $p_a^2 = -s$ and $p_b^2 = -u$, where 
$s$ and $u$ are the invariant Mandelstam variables, have been used. 
These results for 
the $A$ and $B$ amplitudes are formally equivalent to the
corresponding results for the two-pion emission
amplitude for nucleons~\cite{Hamilton}. Note that in 
eq.~(\ref{B-}), the contribution from the Weinberg-Tomozawa interaction 
tends to cancel the constant term in the $B^-$ amplitude that 
arises from the Born terms. If the axial coupling constant $g_A^q$ is 
taken to equal 1, then this cancellation is exact.

\subsection{The Width for Two-Pion Decay}

When excited $D$ mesons decay by emission of pions, it is 
only the light constituent quark component that couples to the pions.
Although pion decay from a single non-interacting quark is 
kinematically impossible, the pion emission amplitude nevertheless
involves the charm antiquark indirectly through the bound state
wave function in the matrix element of the single quark
pion emission operator. The role of two quark emission operators
that involve the interaction between the light quark and
charm antiquark, with pion emission from the former, will be 
considered below in section~\ref{Twosec}. 

The general expression for the two pion decay width of an excited $D$ 
meson may be written in the form:

\begin{equation}
\Gamma = (2\pi)^4\int\frac{d^3k_a}{(2\pi)^3}\frac{d^3k_b}{(2\pi)^3}
\frac{d^3P_f}{(2\pi)^3} \frac{|T_{fi}|^2}
{4\omega_a\omega_b} \delta^{(4)}(P_f + k_a 
+ k_b -P_i).
\end{equation}
Here $k_a$ and $k_b$ denote the four-momenta of the two emitted pions,
$P_i$ and $P_f$ denote those of the initial and final state $D$ mesons 
while $\omega_a$ and $\omega_b$ denote the energies of the 
emitted pions. Since the constituents of the $D$ mesons form bound 
states, their normalization factors are included in the spinors $\bar 
u(p')$ and $u(p)$ in eq.~(\ref{ampl}). In the laboratory frame 
$P_i^0 = M_i$. By introducing the variables $\vec Q = (\vec k_b - \vec 
k_a)/2$ and $\vec q = \vec k_b + \vec k_a$, the decay 
width expression may be rewritten as

\begin{equation}
\Gamma = \int\frac{d^3q d^3Q}{(2\pi)^5} \frac{|T_{fi}|^2}
{4\omega_a\omega_b}\:\:
\delta \left(\sqrt{q^2 + M_f^2} + \omega_a + \omega_b - M_i\right).
\end{equation}
Here the energy factors are defined as 
$\omega_a = \sqrt{m_\pi^2 + (\vec q/2 
-\vec Q)^2}$ and $\omega_b = \sqrt{m_\pi^2 + (\vec q/2 +\vec Q)^2}$ 
respectively. The remaining delta function may be used to fix the 
variable $Q$, so that finally the expression for the differential 
width becomes

\begin{equation}
\frac{d\Gamma}{d\Omega_q} = \frac{1}{4}\frac{1}{(2\pi)^4} \int_0^{q_f}
dq\:q^2 \int_{-1}^1 dz\:\frac{Q_f^2(q,z)}{\omega_a(q,z)\left(Q_f +
\frac{qz}{2}\right) + \omega_b(q,z)
\left(Q_f - \frac{qz}{2}\right)}\:|T_{fi}|^2.
\label{dec}
\end{equation}
Here it is understood that in order to obtain the total widths for 
two-pion decay, eq.~(\ref{dec}) is to be integrated over $\Omega_q$ 
yielding an additional factor $4\pi$. In eq.~(\ref{dec}), the variable $z$ 
is defined by $\vec Q\cdot\vec q = Qqz$. With this notation the pion 
energies $\omega_a$ and $\omega_b$ are given by the expressions $\omega_a 
= \sqrt{m_{\pi}^2 + Q_f^2 + q^2/4 - Q_fqz}$ and
$\omega_b = \sqrt{m_{\pi}^2 + Q_f^2 + q^2/4 + Q_fqz}$, where the fixed 
variable $Q_f$ is given by

\begin{equation}
Q_f^2 = \frac{(E_f - M_i)^4 - (4m_{\pi}^2 + q^2)(E_f - M_i)^2}{4(E_f -
M_i)^2 - 4q^2z^2}.
\end{equation}
Here $E_f$ denotes the energy of the final state $D$ meson and is given 
by $E_f = \sqrt{q^2 + M_f^2}$. In all these expressions, $m_\pi$ denotes 
the pion mass. Since different charge states are not considered in this 
paper, an average pion mass of 137 MeV has been used throughout the 
calculations. The cutoff momentum $q_f$ corresponds to the maximal 
momentum of any one of the final state particles, e.g. the final state $D$ 
meson. Thus $q_f$ corresponds to the q-value of a decay of the form $D' 
\rightarrow DX$, where $D'$ and $D$ are the appropriate $D$ meson states 
and $X$ is a particle with mass $M_X = 2m_\pi$. The appropriate values for 
$q_f$ are listed along with the employed $D$ meson masses in 
Table~\ref{masstab}.

The decomposition in eq.~(\ref{plmi}) is convenient with respect to 
the summation over the isospins of the pions, leading to

\begin{equation}
\sum_{\mathrm{isospin}}|T_{fi}|^2 = 3T^{+\dag} T^+ + 6T^{-\dag} T^-.
\label{isum}
\end{equation}
For calculational purposes, it is useful to split eq.~(\ref{ampl}) into 
spin independent and spin-dependent parts as

\begin{equation}
T^{\pm} = \alpha^{\pm} + i\vec\sigma_q\cdot\vec\beta^{\pm}.
\label{ssum}
\end{equation}
Here, the amplitude
$\alpha$ will contain, in addition to the spin independent amplitude 
$A$, the spin-independent contributions that arise from the $\gamma\cdot 
Q$ term in eq.~(\ref{ampl}), while $\vec \beta$ will contain 
all the spin-dependent terms contained in the expression eq.~(\ref{ampl}). 
Consider first the non-relativistic limit. 
In that limit, the expressions for $\alpha$ and $\vec\beta$ become

\begin{eqnarray}
\alpha^\pm &=& A^\pm + \left(Q_0 -\frac{\vec P\cdot\vec 
Q}{m_q}\right)B^\pm, \label{alpha}\\
\vec\beta^\pm &=& \frac{\vec q\times\vec Q}{2m_q} B^\pm, \label{beta}
\end{eqnarray}
where $Q_0$ is defined as $(\omega_b - \omega_a)/2$ and $\vec P$ is 
defined as $\vec P = 
\left(\vec {p}\,'_q + \vec p_q\right)/2$. 
Because two pion decays of charm mesons with $L=1$ to the ground states
with $L=0$ are considered here, the largest contribution is expected to 
arise from the $\vec P$ dependent term in the spin independent amplitude 
$\alpha^\pm$.

Evaluation of the 
expression~(\ref{isum}) requires spin sums for both the 
spin independent and spin-dependent terms in eq.~(\ref{ssum}). 
The spin independent part of the amplitude requires evaluation
of the spin sums

\begin{equation}
\frac{1}{2J+1} \sum_{M=-J}^{J} \left<LSJM\right|\alpha^*
\left(\frac{1-\vec\sigma_q\cdot\vec\sigma_{\bar Q}}{4}\right) \alpha 
\left|LSJM\right>,
\label{sing}
\end{equation}

\begin{equation}
\frac{1}{2J+1} \sum_{M=-J}^{J} \left<LSJM\right|\alpha^*
\left(\frac{3+\vec\sigma_q\cdot\vec\sigma_{\bar Q}}{4}\right) \alpha 
\left|LSJM\right>,
\label{trip}
\end{equation}  
for decay to the spin singlet $D$ and spin triplet $D^*$ mesons
respectively. For 
two-pion decay of the spin triplet $D_1(2420)$ and $D_2^*(2460)$
mesons the spin independent amplitude gives no contribution for decay
to the ground state $D$ meson (\ref{sing}).
The contribution given by the expressions above
is $|\alpha|^2$ 
to the decay rate to the spin triplet $D^*$ meson (\ref{trip}). 

The corresponding spin sums for the 
$\vec\sigma_q\cdot\vec\beta$ dependent term in eq.~(\ref{ssum}) are 
somewhat more complicated. For initial states with 
$L=1,S=1$ they may be expressed as in ref.~\cite{TL3}, giving

\begin{equation}
\frac{1}{2J+1} \sum_{M=-J}^{J} \left<11JM\right|\frac{\vec\beta^2}{3}
-\frac{S_{12}(\vec\beta)}{6}\left|11JM\right>,
\label{sum3}
\end{equation} 
for decay to spin singlet ($D$) final states, and

\begin{equation}
\frac{1}{2J+1} \sum_{M=-J}^{J} \left<11JM\right|\frac{2\vec\beta^2}{3}  
+\frac{S_{12}(\vec\beta)}{6}\left|11JM\right>,
\label{sum4}
\end{equation}  
for decay to spin triplet ($D^*$) final states. In eqs.~(\ref{sum3}) 
and~(\ref{sum4}), $S_{12}(\hat\beta)$ is defined as the 
tensor operator 
$S_{12}(\hat\beta) = 3\vec\sigma_q\cdot\hat\beta
\vec\sigma_{\bar Q}\cdot\hat\beta - \vec\sigma_q\cdot\vec\sigma_{\bar Q}$. 
Given these spin sums, and the state vectors for the
spin 1 $P$-state $D$ mesons,

\begin{equation}
\left|1SJM\right>=\sum_{ls}\left(11 ls|JM\right){u_1(r)\over r}Y_{1l}(\hat 
r)\left|1s\right> + \sum_{l}\left(10 l0|JM\right){u_1(r)\over  r}Y_{1l}(\hat
r)\left|00\right>,
\end{equation}
where $\left|1s\right>$ denotes a spin triplet state with $s_z = s$,
the following spin summed squared amplitudes for the two-pion decays 
of the spin triplet $D_2^*$, $D_1$ and $D_0^*$ mesons are obtained:

\begin{eqnarray}
|T|_{D_2^* \rightarrow D^*}^2 &=& 3\left[|\alpha^+|^2 + 2|\alpha^-|^2 + 
\frac{3}{5}\left(|\vec\beta^+|^2 + 
2|\vec\beta^-|^2\right)\right], \label{Ddec2} \\
|T|_{D_1 \rightarrow D^*}^2 &=& 3\left[|\alpha^+|^2 + 2|\alpha^-|^2 + 
|\vec\beta^+|^2 + 2|\vec\beta^-|^2 \right], \label{D1dec} \\
|T|_{D_0^* \rightarrow D^*}^2 &=& 3\left[ |\alpha^+|^2 + 
2|\alpha^-|^2 \right], \label{D0dec2} \\
|T|_{D_2^* \rightarrow D}^2 &=& 3\left[ \frac{2}{5}\left(|\vec\beta^+|^2 + 
2|\vec\beta^-|^2\right)\right], \label{Ddec1} \\
|T|_{D_1 \rightarrow D}^2 &=& 0, \\
|T|_{D_0^* \rightarrow D}^2 &=& 3\left[ |\vec\beta^+|^2 + 
2|\vec\beta^-|^2 \right]. \label{D0dec1}
\end{eqnarray}
From the above results, it follows that for the $\pi\pi$ decays of the 
spin triplet $D_1$ meson, there is no contribution for two-pion decay 
to the $D$ states. For the spin singlet $D_1^*$ meson, similar expressions 
hold, with the modification that the spin-independent amplitudes now only 
contribute to decay to $D$ states. For the spin-dependent terms, one has 
to consider the following additional spin sums

\begin{equation}
\frac{1}{3} \sum_{M=-1}^{1} \left<101M\right|
-\frac{S_{12}(\vec\beta)}{6}\left|101M\right>,
\label{sum5}
\end{equation}  
for decay to spin singlet $D$ final states, and

\begin{equation}
\frac{1}{3} \sum_{M=-1}^{1} \left<101M\right|\vec\beta^2  
+\frac{S_{12}(\vec\beta)}{6}\left|101M\right>,
\label{sum6}
\end{equation}  
for decay to spin triplet $D^*$ final states. Application of these spin 
sums yields then the desired expressions for the spin summed squared 
amplitudes for two-pion decay of the spin singlet $D_1^*$ meson:

\begin{eqnarray}
|T|_{D_1^* \rightarrow D^*}^2 &=& 3\left[ |\vec\beta^+|^2 + 
2|\vec\beta^-|^2 \right], \label{D1*dec1} \\
|T|_{D_1^* \rightarrow D}^2 &=& 3\left[ |\alpha^+|^2 + 
2|\alpha^-|^2 \right]. \label{D1*dec2}
\end{eqnarray}
In the non-relativistic limit, the following two radial matrix elements are 
required for the numerical evaluation of the above amplitudes:

\begin{eqnarray}
{\cal M}_0 &=& \frac{1}{3m_q}\int_{0}^{\infty}dr
\left[u'_0(r)u_1(r)-u_0(r)u'_1(r)-2{u_0(r)u_1(r)\over 
r}\right]\,j_0\left({qr\over 2}\right), \label{m0} \\
{\cal M}_1 &=& \int_0^{\infty} dr\:
u_0(r)u_1(r)\:j_1\left(\frac{qr}{2}\right). \label{m1}
\end{eqnarray}
Here $u_0$ and $u_1$ denote the reduced radial wavefunctions 
that are obtained by numerical solution of the Blankenbecler-Sugar 
equation in ref.~\cite{TL2} for states with $L=0,1$ respectively.

In the following calculation, the non-local
combinations of the kinematic variables, which appear
in the denominators in the Born term amplitudes $B^+$ and $B^-$, have been 
approximated by their corresponding expectation values. For calculational 
convenience, the approximation $E_q \simeq m_q$ has likewise been made. 
The kinematical variables $\vec Q$ and $\vec q$ have, in addition to the 
energies of the emitted pions, been treated without approximation. The 
resulting expressions are then obtained as 

\begin{eqnarray}
B^+ &\simeq& -4m_q^2 \left(\frac{g_A^q}{2f_\pi}\right)^2
\left\{\left[m_\pi^2 + \left<\vec P\cdot\vec q\right> 
-2\left<\vec P\cdot\vec Q\right> + \frac{q^2}{2} - \vec Q\cdot\vec q
-2m_q\omega_a \right]^{-1} \right. \nonumber \\
&&\left.-\left[m_\pi^2 + \left<\vec P\cdot\vec q\right>     
+2\left<\vec P\cdot\vec Q\right> + \frac{q^2}{2} + \vec Q\cdot\vec q
-2m_q\omega_b \right]^{-1} \right\}, \label{Bamp+} \\
B^- &\simeq& -4m_q^2 \left(\frac{g_A^q}{2f_\pi}\right)^2
\left\{\left[m_\pi^2 + \left<\vec P\cdot\vec q\right>
-2\left<\vec P\cdot\vec Q\right> + \frac{q^2}{2} - \vec Q\cdot\vec q
-2m_q\omega_a \right]^{-1} \right. \nonumber \\
&&\left.+\left[m_\pi^2 + \left<\vec P\cdot\vec q\right>
+2\left<\vec P\cdot\vec Q\right> + \frac{q^2}{2} + \vec Q\cdot\vec q
-2m_q\omega_b \right]^{-1} \right\} +\frac{1 - g_A^{q\:2}}{2f_\pi^2}.
\label{Bamp-}
\end{eqnarray}
The contribution to the decay rate from the spin independent
$\alpha^+$ amplitude in (\ref{ssum}) may then be expressed as

\begin{equation}
|T|_{\alpha^+}^2 = 3|\alpha^+|^2 = 
3\left[A^+{\cal M}_1 + B^+\left(\frac{\omega_b - \omega_a}{2}
{\cal M}_1 - \frac{Q_fz}{2}{\cal M}_0 \right)\right]^2
\label{alpha+}
\end{equation}  
for all decays to $D^*$ final states. Note that the matrix elements here 
are $q$-dependent. Similarly, by using the 
$B^-$ amplitude, the contribution to $|T|^2$ from the $\alpha^-$ term 
becomes

\begin{equation}
|T|_{\alpha^-}^2 = 6|\alpha^-|^2 = 
6\left(B^-\right)^2\left[\frac{\omega_b - \omega_a}{2}
{\cal M}_1 - \frac{Q_fz}{2}{\cal M}_0 \right]^2,
\label{alpha-}
\end{equation}
for all decays to $D^*$ final states. Here $\omega_a$ and 
$\omega_b$ are defined as for eq.~(\ref{dec}). In addition, the $B^+$ 
and $B^-$ amplitudes also contribute through the spin dependent amplitudes
$\vec\beta^+$ and $\vec\beta^-$~(\ref{ssum}).
In the case of the decay mode $D_2^* \rightarrow D^*\pi\pi$, these 
contributions may be expressed as

\begin{equation}
|T|_{\beta^-}^2 = \frac{18}{5}|\vec\beta^-|^2 = 
\frac{3}{2}\frac{q^2Q_f^2(1-z^2)}{m_q^2} \left(B^-\right)^2 \frac{3}{5}
{\cal M}_1^2,
\label{beta-}
\end{equation}
and

\begin{equation}
|T|_{\beta^+}^2 = \frac{9}{5}|\vec\beta^+|^2 = 
\frac{3}{4}\frac{q^2Q_f^2(1-z^2)}{m_q^2} \left(B^+\right)^2 \frac{3}{5}
{\cal M}_1^2.
\label{beta+}
\end{equation}
The corresponding expressions for the remaining decay modes of the spin 
triplet $L=1$ $D$ mesons to which the spin-dependent amplitudes contribute 
may be inferred from eqs.~(\ref{Ddec2}-\ref{D0dec1}): For the decay 
$D_2^*\rightarrow D\pi\pi$, 
the factors 3/5 on the r.h.s. of eqs.~(\ref{beta-}) and~(\ref{beta+}) are 
to be replaced by 2/5, and for the decays $D_0^*\rightarrow D\pi\pi$ and
$D_1\rightarrow D^*\pi\pi$, those factors equal 1. That is also the case 
for the decay of the spin singlet $D_1^*$ meson to $D^*\pi\pi$ 
(\ref{D1*dec2}). The total decay widths are obtained by adding the 
contributions from eqs.~(\ref{alpha+},\ref{alpha-},\ref{beta-}) 
and~(\ref{beta+}), and 
inserting them into eq.~(\ref{dec}). In the numerical evaluation of the 
expressions~(\ref{Bamp+}) and~(\ref{Bamp-}), the terms $\vec P\cdot\vec 
q$ and $\vec P\cdot\vec Q$ in the denominators have been approximated by 
their respective expectation values:

\begin{eqnarray}
\left<\vec P\cdot\vec q\right> &=& \frac{q\,m_q}{2}{\cal M}_0, \\
2\left<\vec P\cdot\vec Q\right> &=& Q_f z\,m_q {\cal M}_0.
\end{eqnarray}
Note that despite the appearance of the matrix element 
${\cal M}_0$ in the above expressions, no non-relativistic approximation 
is implied. 

As the velocities of the confined quarks and antiquarks in the 
$D$ mesons are close to that of light, the non-relativistic approximation 
for the amplitudes is not reliable. 
Therefore, eqs.~(\ref{alpha}) and~(\ref{beta}) should be replaced by the 
analogous unapproximated forms

\begin{eqnarray}
\alpha^\pm &=& \sqrt{{E'+m_q\over 2E'}}\sqrt{{E+m_q\over 2E}}\left[A^\pm
\left(1-{P^2-q^2/4\over (E'+m_q)(E+m_q)}\right) + B^\pm \left\{Q_0
\left(1+{P^2-q^2/4\over (E'+m_q)(E+m_q)}\right) \right.\right. 
\nonumber \\
&& \left.\left. -\vec P\cdot\vec Q 
\left(\frac{1}{E+m_q}+\frac{1}{E'+m_q}\right)
-\frac{\vec Q\cdot\vec q}{2}\left(\frac{1}{E+m_q}-\frac{1}{E'+m_q}\right)
\right\}\right],
\label{alpharel}
\end{eqnarray}
and

\begin{eqnarray}
\vec\beta^\pm &=& \sqrt{{E'+m_q\over 2E'}}\sqrt{{E+m_q\over 
2E}}\left\{\frac{A^\pm - B^\pm Q_0}{(E+m_q)(E'+m_q)} \vec q\times\vec P + 
B^\pm\left[\vec P\times\vec Q \left(\frac{1}{E+m_q} - 
\frac{1}{E'+m_q}\right) 
\right.\right. \nonumber \\
&& \left.\left. + \frac{\vec q\times\vec Q}{2}\left(\frac{1}{E+m_q} + 
\frac{1}{E'+m_q}\right)\right]\right\}. \label{betarel}
\end{eqnarray}
Thus, relativistic counterparts to eqs.~(\ref{alpha+}-\ref{beta+}) are 
required. As a consequence, eqs.~(\ref{alpha+}) 
and~(\ref{alpha-}) should be replaced by the expressions 

\begin{equation}
|T|_{\alpha^+}^2 = 3|\alpha^+|^2 = 
3\left[A^+{\cal M}_{1+}^{\mathrm{rel}}
+ B^+\left(\frac{\omega_b - \omega_a}{2} {\cal M}_{1-}^{\mathrm{rel}}
- \frac{Q_fz}{2}{\cal M}_{0}^{\mathrm{rel}}\right)\right]^2,
\label{ny1}
\end{equation}  
and

\begin{equation}
|T|_{\alpha^-}^2 = 6|\alpha^-|^2 = 
6\left(B^-\right)^2\left[\frac{\omega_b - \omega_a}{2}
{\cal M}_{1-}^{\mathrm{rel}} - \frac{Q_fz}{2}
{\cal M}_0^{\mathrm{rel}}\right]^2.
\label{ny2}
\end{equation}
Here the relativistic matrix elements ${\cal M}_1^{\mathrm{rel}}$ are 
defined as

\begin{eqnarray}
{\cal M}_{1\pm}^{\mathrm{rel}}&=&{1\over
\pi}\int_{0}^{\infty}dr'r'u_0(r')\int_{0}^{\infty}dr\,r\,
u_1(r)\int_{0}^{\infty}dP\,P^2\int_{-1}^{1}dv\,f_{\mathrm{BS}}(P,v)
\nonumber \\
&& {q/4+Pv\over \sqrt{P^2+q^2/16+Pqv/2}}
\:\sqrt{{E'+m_q\over 2E'}}\sqrt{{E+m_q\over 2E}}
\left(1\mp{P^2-q^2/4\over (E'+m_q)(E+m_q)}\right)
\nonumber \\
&&j_0\left(r'\sqrt{P^2+{q^2\over 16}-{Pqv\over
2}}\:\right)j_1\left(r\sqrt{P^2+{q^2\over
16}+{Pqv\over 2}}\:\right), \label{M1rel} 
\end{eqnarray}
where the last term in eq.~(\ref{alpharel}) has been neglected because of 
its smallness. Similarly, the relativistic matrix element 
${\cal M}_0^{\mathrm{rel}}$ 
is obtained as

\begin{eqnarray}
{\cal M}_0^{\mathrm{rel}}&=&{1\over
3\pi}\int_{0}^{\infty}dr'r'\int_{0}^{\infty}dr\,r\int_{0}^{\infty}
dP\,P^2 \int_{-1}^{1}dv\,f_{\mathrm{BS}}(P,v) 
\nonumber \\
&&
\sqrt{{E'+m_q\over 2E'}}\sqrt{{E+m_q\over 2E}}
\left(\frac{1}{E+m_q}+\frac{1}{E'+m_q}\right) 
\left[u'_0(r')u_1(r)-u_0(r')u'_1(r)-2{u_0(r')u_1(r)\over r}\right]
\nonumber \\
&&j_0\left(r'\sqrt{P^2+{q^2\over
16}-{Pqv\over 2}}\:\right)j_0\left(r\sqrt{P^2+{q^2\over 16}+{Pqv\over
2}}\:\right). 
\label{M2rel}
\end{eqnarray}
In the non-relativistic limit, these matrix elements reduce to the forms of 
eqs.~(\ref{m1}) and~(\ref{m0}) respectively. In the above equations, the
factor $f_{\mathrm{BS}}(P,v)$ arises in the reduction of the amplitude
from the form appropriate to the Bethe-Salpeter equation to that of the 
Blankenbecler-Sugar equation, and is defined as

\begin{equation}
f_{\mathrm{BS}}(P,v)=\frac{M_Q+m_q}{\sqrt{(E+E_c)(E'+E'_c)}}.
\label{Blsu}
\end{equation}
Here $M_Q$ denotes the mass of the heavy (anti)quark. The energy factors 
$E$,$E'$ and $E_c$,$E'_c$ are defined as

\begin{equation}
E=\sqrt{m_q^2+P^2+Pqv+q^2/4},\qquad
E'=\sqrt{m_q^2+P^2-Pqv+q^2/4},
\end{equation}
for the light quark $q$, and

\begin{equation}
E_c=\sqrt{M_Q^2+P^2+Pqv+q^2/4},\qquad
E'_c=\sqrt{M_Q^2+P^2-Pqv+q^2/4},
\end{equation}
for the heavy (anti)quark $Q$. In addition, 
eqs.~(\ref{beta-}) and~(\ref{beta+}) should also be replaced by analogous 
relativistic forms. These may be obtained as

\begin{equation}
|T|_{\beta^-}^2 = 6 \left(B^-\right)^2 \frac{3}{5} 
\left[\frac{q\,Q_f\sqrt{1-z^2}}{2}\:{\cal M}_{\beta 1}^{\mathrm{rel}}
-\frac{\omega_b - \omega_a}{2}\,q\:{\cal M}_{\beta 
2}^{\mathrm{rel}}\right]^2,
\label{betar-}
\end{equation}
for the contribution from $\vec\beta^-$, and

\begin{equation}
|T|_{\beta^+}^2 = 3\:\frac{3}{5} \left\{A^+q{\cal M}_{\beta
2}^{\mathrm{rel}} + B^+
\left[\frac{q\,Q_f\sqrt{1-z^2}}{2}\:{\cal M}_{\beta 1}^{\mathrm{rel}}
-\frac{\omega_b - \omega_a}{2}\,q\:{\cal M}_{\beta 
2}^{\mathrm{rel}}\right]\right\}^2,
\label{betar+}
\end{equation}
for the corresponding one from $\vec\beta^+$. Here the matrix elements 
are defined as

\begin{eqnarray}
{\cal M}_{\beta 1}^{\mathrm{rel}}&=&{1\over\pi}
\int_{0}^{\infty}dr'r'u_0(r')\int_{0}^{\infty}dr\,r\,
u_1(r)\int_{0}^{\infty}dP\,P^2\int_{-1}^{1}dv\,f_{\mathrm{BS}}(P,v)
\nonumber \\
&& {q/4+Pv\over \sqrt{P^2+q^2/16+Pqv/2}}
\:\sqrt{{E'+m_q\over 2E'}}\sqrt{{E+m_q\over 2E}}
\left({1\over E'+m_q}+{1\over E+m_q}\right)
\nonumber \\
&&j_0\left(r'\sqrt{P^2+{q^2\over 16}-{Pqv\over
2}}\:\right)j_1\left(r\sqrt{P^2+{q^2\over
16}+{Pqv\over 2}}\:\right), 
\label{M1'rel} 
\end{eqnarray}
and

\begin{eqnarray}
{\cal M}_{\beta 2}^{\mathrm{rel}}&=&{1\over\pi}
\int_{0}^{\infty}dr'r'u_0(r')\int_{0}^{\infty}dr\,r\,
u_1(r)\int_{0}^{\infty}dP\,P^3\int_{-1}^{1}dv\,\sqrt{1-v^2}\,
f_{\mathrm{BS}}(P,v) \nonumber \\
&& {q/4+Pv\over \sqrt{P^2+q^2/16+Pqv/2}}
\:\sqrt{{E'+m_q\over 2E'}}\sqrt{{E+m_q\over 2E}}
{1\over (E'+m_q)(E+m_q)}
\nonumber \\
&&j_0\left(r'\sqrt{P^2+{q^2\over 16}-{Pqv\over
2}}\:\right)j_1\left(r\sqrt{P^2+{q^2\over
16}+{Pqv\over 2}}\:\right).
\label{Mbeta2rel} 
\end{eqnarray}
With exception of a factor $m_q^{-1}$, the matrix element~(\ref{M1'rel}) 
reduces to the form~(\ref{m1}) in the non-relativistic limit. In these 
expressions, the $\vec P \times \vec Q$ term in eq.~(\ref{betarel}) has 
been dropped because of its smallness. Furthermore, it turns out that 
eq.~(\ref{Mbeta2rel}) is numerically much smaller than eq.~(\ref{M1'rel}). 
Overall, the contributions from 
the amplitudes $\alpha^+$ and $\alpha^-$ are dominant, whereas 
those from $\vec\beta^+$ and $\vec\beta^-$ represent only small 
corrections. This feature is in accordance with the current experimental 
status for the analogous decays of the excited strange mesons, for
which the decay mode $K_2^*\rightarrow K\pi\pi$ 
has not yet been seen, although the decays of the $K_2^*$ meson 
are otherwise well studied. In eqs.~(\ref{betar-}) and~(\ref{betar+}), it 
is again understood that the factor 3/5 is to be replaced as for 
eqs.~(\ref{beta-}) and~(\ref{beta+}) when different decay modes are 
considered. The constituent quark and $D$ meson masses used are listed in 
Table~\ref{masstab}. The numerical results obtained by using the 
relativistic expressions are given in Table~\ref{reltab}.
For comparison the corresponding non-relativistic results are presented in 
Table~\ref{nrtab}. The non-relativistic treatment leads to overpredictions 
by a factor $\sim 3$.

\newpage

\begin{table}[h!]
\begin{center}
\begin{tabular}{l|c|c|c}
\quad Decay & Initial state mass ($M_i$) & Final state mass ($M_f$) & 
$q_f$ \\ \hline\hline &&&\\
$D_2^*\rightarrow D^*\pi\pi$ & 2460 & 2007 & 327 \\
$D_2^*\rightarrow D \pi\pi$  & 2460 & 1867 & 462 \\
$D_1  \rightarrow D^*\pi\pi$ & 2420 & 2007 & 282 \\
$D_1^*\rightarrow D^*\pi\pi$ & 2389 & 2007 & 244 \\ 
$D_1^*\rightarrow D \pi\pi$  & 2389 & 1867 & 395 \\ 
$D_0^*\rightarrow D^*\pi\pi$ & 2341 & 2007 & 177 \\ 
$D_0^*\rightarrow D \pi\pi$  & 2341 & 1867 & 347 \\ &&&
\end{tabular}
\caption{Initial and final state $D$ meson masses and resulting maximal 
kinematically allowed momenta $q_f$ in MeV used in the calculations. 
The masses listed are to be viewed as averages over the different 
charge states. In the calculations, the values $m_q$ = 450 MeV and 
$M_Q$ = 1580 MeV, obtained by fits to the spectra in ref.~\cite{TL2}, 
have been used  
for the constituent quark masses.} \label{masstab}
\end{center}
\end{table}

\begin{table}[h!]
\begin{center}
\begin{tabular}{l|c|c|c|c|c|c}
\quad Decay & $|T|_{\alpha^+}^2$ & $|T|_{\alpha^-}^2$ & $|T|_{\beta^-}^2$ 
& $|T|_{\beta^+}^2$ & Total ($g_A^q=1$) & Total ($g_A^q=0.87$)
\\ \hline\hline &&&&&&\\
$D_2^*\rightarrow D^*\pi\pi$ & 0.896 & 2.864 & $5.06\cdot 10^{-3}$ &
$1.17\cdot 10^{-3}$ & 3.77 MeV & 2.39 MeV \\
$D_2^*\rightarrow D \pi\pi$  & --    & --    & $6.45\cdot 10^{-2}$ &
$7.51\cdot 10^{-3}$ & 0.07 MeV & 0.05 MeV \\
$D_1  \rightarrow D^*\pi\pi$ & 0.367 & 1.291 & $2.33\cdot 10^{-3}$ &
$7.85\cdot 10^{-4}$ & 1.66 MeV & 1.05 MeV \\ 
$D_1^*\rightarrow D^*\pi\pi$ & --    & --    & $6.54\cdot 10^{-4}$ &
$3.30\cdot 10^{-4}$ & $\simeq 0$ MeV & $\simeq 0$ MeV \\
$D_1^*\rightarrow D \pi\pi$  & 2.749 & 7.915 & --                  &
--                  & 10.7 MeV & 6.80 MeV \\
$D_0^*\rightarrow D^*\pi\pi$ & 0.020 & 0.098    
& -- & --  & 0.12 MeV & 0.07 MeV \\
$D_0^*\rightarrow D \pi\pi$ & --    & --    & $1.43\cdot 10^{-2}$ &
$2.86\cdot 10^{-3}$ & 0.02 MeV & 0.01 MeV \\ &&&&&&
\end{tabular}
\caption{Numerical results using the single-quark amplitudes and the 
relativistic matrix elements for the 
two-pion decay widths of the spin triplet $D_2^*$, $D_1$, $D_0^*$ and the 
spin singlet $D_1^*$ mesons. The individual results from 
the spin independent ($\alpha^\pm$) and spin dependent amplitudes 
($\beta^\pm$) are shown for $g_A^q$ = 1, and the total 
$\pi\pi$ decay widths for both 
$g_A^q$ = 1 and $g_A^q$ = 0.87.} \label{reltab}
\end{center}
\end{table}

The results for the two-pion decay width of the positive
parity charm mesons with $L=1$ shown in Table~\ref{reltab}
reveal a strong sensitivity to the strength of the pion coupling to the 
light constituent quarks as measured by the axial coupling constant of the
constituent quarks. This is of course as expected, as the
calculated decay widths are proportional to the 4th power
of $g_A^q$. Also, because of the large~(140 MeV) splitting between the $D$ 
and $D^*$ states and the significant spin-orbit splittings of the $L=1$ 
charm mesons, it turns out that the values of 
the maximal kinematically allowed momentum transfer
$q_f$ as given in 
Table~\ref{masstab} show marked variation. Consequently, some decays are 
kinematically favored, while others, in particular $D_0^*\rightarrow 
D^*\pi\pi$, are strongly inhibited by the small phase space available. 
Hence the two-pion decay widths of the $L=1$ $D$ 
mesons are very sensitive to the spin-orbit structure of the 
quark-antiquark interaction. The same conclusion was also reached in 
ref.~\cite{TL3} concerning the single pion decays of the $L=1$ $D$ mesons. 
In the absence of empirical information and definite QCD lattice
calculations~\cite{Wolo1,Wolo2,Khan,Boyle}, the energies of the hitherto 
undiscovered $D_0^*$ and $D_1^*$ mesons have here been taken to equal 
those found in the calculation of ref.~\cite{TL3}.

The predicted total decay widths of the $L=1$ $D$ mesons in the single 
quark approximation may be obtained by adding the calculated 
two-pion decay widths in Table~\ref{reltab} to those for single pion decay 
obtained in ref.~\cite{TL2}. With $g_A^q=1$ the total 
calculated $\pi\pi$ decay width of the $D_2^*(2460)$ is 3.8 MeV and that 
of the $D_1(2420)$ is 1.7 MeV. If these values are added to the 
corresponding calculated values for single pion decay obtained in 
ref.~\cite{TL3}, the total calculated width of the $D_2^*(2460)$ comes to
19.5 MeV, which is well within the uncertainty margin of
the empirical value $25_{-7}^{+8}$ MeV for the
total decay width~\cite{PDG}. In the case of the $D_1(2420)$ meson the 
total calculated width for single and two-pion decay comes
to 15.3 MeV, which is close to the empirical uncertainty margin of the 
total decay width $18.9_{-3.5}^{+4.6}$ MeV~\cite{PDG}. Reduction of the 
value for $g_A^q$ to 0.87 brings the calculated values for the total width 
for $\pi$ and $\pi\pi$ decay a bit below the empirical values for the total
widths. Likewise, employment of the two-quark contribution considered in 
the next section has the effect of reducing the calculated $\pi\pi$ 
widths. The final results for the $\pi\pi$ widths are listed in 
Table~\ref{tottab}, along with predictions for the total widths of the 
$L=1$ $D$ mesons.

\vspace{0.7cm}

\begin{table}[h!]
\begin{center}
\begin{tabular}{l|c|c|c|c|c|c}
\quad Decay & $|T|_{\alpha^+}^2$ & $|T|_{\alpha^-}^2$ & $|T|_{\beta^-}^2$ 
& $|T|_{\beta^+}^2$ & Total ($g_A^q=1$) & Total ($g_A^q=0.87$)
\\ \hline\hline &&&&&&\\
$D_2^*\rightarrow D^*\pi\pi$ & 2.638 & 9.934 & $1.36\cdot 10^{-2}$ &
$3.11\cdot 10^{-4}$ & 12.6 MeV & 8.01 MeV \\
$D_2^*\rightarrow D \pi\pi$  & --    & --    & 0.178 &
$5.77\cdot 10^{-3}$ & 0.18 MeV & 0.12 MeV \\
$D_1  \rightarrow D^*\pi\pi$ & 1.083 & 4.539 & $6.21\cdot 10^{-3}$ &
$1.16\cdot 10^{-4}$ & 5.63 MeV & 3.56 MeV \\
$D_1^*\rightarrow D^*\pi\pi$ & --    & --    & $1.74\cdot 10^{-3}$ &
$2.60\cdot 10^{-5}$ & $\simeq 0$ MeV & $\simeq 0$ MeV \\
$D_1^*\rightarrow D  \pi\pi$ & 8.048 & 26.97 & -- &
--                  & 35.0 MeV & 22.4 MeV \\
$D_0^*\rightarrow D^*\pi\pi$ & 0.059 & 0.357 & -- 
& -- & 0.42 MeV & 0.26 MeV \\
$D_0^*\rightarrow D  \pi\pi$ & --    & --    & $3.84\cdot 10^{-2}$ &
$9.51\cdot 10^{-4}$ & 0.04 MeV & 0.03 MeV \\ &&&&&&
\end{tabular}
\caption{Numerical results for the two-pion decay
widths of the spin 
triplet $D_2^*$, $D_1$, $D_0^*$ and the spin singlet $D_1^*$ mesons using 
the non-relativistic approximation to the single-quark amplitudes. The 
individual results from $\alpha^\pm$ and $\beta^\pm$ are 
shown for $g_A^q$ = 1, and the total decay widths for both $g_A^q$ = 1 and 
$g_A^q$ = 0.87.} \label{nrtab}
\end{center}
\end{table}

This apparent slight underprediction of the total decay widths may 
actually be a desirable situation in view of the fact that it is 
kinematically possible for other decay modes, mainly $\pi\pi\pi$ and 
$\eta$-meson decay, to contribute to the total widths of these mesons, 
even though there is very little phase space available for such decays. 
However, in view of the large systematical 
uncertainties involved in the experimental determination of the total 
widths of the $D_2^*$ and $D_1$ mesons, the uncertainty margins quoted by 
ref.~\cite{PDG} may be on the narrow side.

\newpage

\section{Two-Quark Contribution to the Decay Width} 
\label{Twosec}

It is instructive, for the purpose of determining the two-quark 
contribution to the decay width, to rewrite the Weinberg-Tomozawa 
Lagrangian, eq.~(\ref{Tomoz}) in the form of a current-current coupling:

\begin{equation}
{\cal L}_\mathrm{WT}=-{1\over 4 f_\pi^2}\vec V_\mu\cdot
\vec\phi_\pi\times\partial_\mu\vec\phi_\pi.
\label{ex1}
\end{equation}
Here $\vec V_\mu = i\bar\psi_q\,\gamma_{\mu}\,\psi_q\,\vec\tau $ is the 
isovector current of the light constituent quark and $\vec\phi_\pi 
\times\partial_\mu\vec\phi_\pi$ is the 
current of the two-pion system. Given this expression it becomes
very natural to describe the irreducible two-quark contribution to the 
two-pion production operator by means of two-quark interaction current 
contributions to the isovector current $\vec V_\mu$.

\begin{figure}[h!]
\begin{center}
\begin{tabular}{c c c} \\
\begin{fmffile}{2piZgraph}
\begin{fmfgraph*}(130,150) \fmfpen{thin}
\fmfleft{i2,i4}
\fmftop{i3,o2}
\fmfbottom{i1,o1}   
\fmf{dashes,label=$\pi$}{i2,v1}
\fmf{dashes,label=$\pi$}{i4,v1}
\fmf{fermion}{i1,v2,v1,i3}
\fmf{dashes,label=$V_c$}{v2,v3}
\fmf{fermion}{o2,v3,o1} 
\fmflabel{$q$}{i1}
\fmfforce{(.25w,.35h)}{v1}
\fmfforce{(.45w,.5h)}{v2}
\fmfforce{(.9w,.5h)}{v3}
\fmfforce{(.35w,0)}{i1}
\fmfforce{(0,.35h)}{i2}   
\fmfforce{(.15w,h)}{i3}
\fmfforce{(0,.60h)}{i4}   
\fmfforce{(w,0)}{o1}   
\fmfv{l=$\bar Q$,l.a=-90,l.d=.04w}{o1}
\fmfforce{(w,h)}{o2}
\end{fmfgraph*}
\end{fmffile}
& \quad \quad \quad \quad &
\begin{fmffile}{2piZgraph2}
\begin{fmfgraph*}(130,150) \fmfpen{thin}
\fmfleft{i2}
\fmftop{i3,o2}
\fmfbottom{i1,o1}   
\fmf{dashes,label=$\pi$}{i2,v1}
\fmf{dashes,label=$\pi$}{i4,v1}
\fmf{fermion}{i1,v1,v2,i3}
\fmf{dashes,label=$V_c$}{v2,v3}
\fmf{fermion}{o2,v3,o1}
\fmflabel{$q$}{i1}
\fmfforce{(.25w,.65h)}{v1}
\fmfforce{(.45w,.5h)}{v2}
\fmfforce{(.9w,.5h)}{v3}
\fmfforce{(.15w,0)}{i1}
\fmfforce{(0,.65h)}{i2} 
\fmfforce{(.35w,h)}{i3}
\fmfforce{(0,.90h)}{i4} 
\fmfforce{(w,0)}{o1}
\fmfv{l=$\bar Q$,l.a=-90,l.d=.04w}{o1}
\fmfforce{(w,h)}{o2}
\end{fmfgraph*}
\end{fmffile}
\\
\\
\\
\end{tabular}

\caption{Two-quark contributions to the pion production amplitude
associated with the Weinberg-Tomozawa Lagrangian. The diagrams shown 
correspond to both time orderings of the two-quark contribution from the 
scalar confining interaction.}
\label{Zdiagram}
\end{center}
\end{figure}
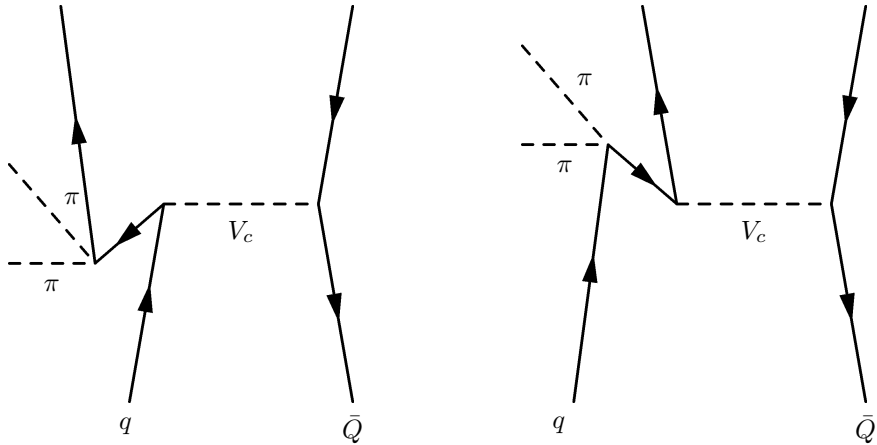

The most significant interaction current contribution to the
isovector current of the charm mesons is the "pair"
current that is associated with the scalar confining
interaction between the constituent quarks. This current,
which is illustrated by the Feynman diagrams in Fig.~\ref{Zdiagram},
represents a relativistic correction term, which arises in the elimination 
of the small components of the quark spinors. In the non-relativistic 
approximation it can be described as a renormalization of the isovector
current of the light constituent quark by the confining
interaction, either in terms of interaction correction
to the constituent quark mass or as an explicit additional
term~\cite{Dannbom}.

In the non-relativistic limit, the isovector current $\vec V_\mu = (i\vec 
V_0,\vec V)$ of the light constituent quark takes the form

\begin{equation}
\vec V={1\over 2 m_q}(\vec p\,'_q +\vec p_q -i\vec \sigma_q
\times\vec q\,)\,\vec \tau,
\label{ex2}
\end{equation}
where $m_q$ is the light constituent quark mass and $\vec p_q$ and   
$\vec p\,'_q$ are the initial and final quark momenta respectively. 
The corresponding expression for the "pair" current that is associated 
with the confining interaction is then obtained as~\cite{Blunden}

\begin{equation}
\vec V_c = -{V_c(k)\over m_q}\vec V.  
\label{ex3}
\end{equation}
Here $V_c(k)$ is the (formal) Fourier transform of the
confining interaction, and the momentum transfer
$\vec k$ is defined as the difference between the final
and initial momenta of the charm (anti)quark:
$\vec k=\vec p\,'_Q -\vec p_Q$. Within the non-relativistic approximation 
it now becomes a straightforward task to take this two-quark contribution 
to the two-pion production operator~\cite{Nyman} into account. The most 
significant contribution may be included by modifying that part of the 
amplitude $B^-$ in the expression for $\alpha^-$~(eq.~(\ref{alpha})), 
which arises from the Weinberg-Tomozawa Lagrangian, according to the 
following prescription:

\begin{equation}
{1\over 2 f_\pi^2}\rightarrow{1\over 2 f_\pi^2}
\left(1-{V_c(r)\over m_q}\right).
\label{ex4}
\end{equation}
Here $V_c(r)$ is the scalar confining interaction, which in the 
wavefunction model of ref.~\cite{TL3} has the form $V_c(r)=cr-b$, with   
$c=1120$ MeV/fm and $b=320$ MeV. This replacement should only be made for 
terms that are associated with the spatial current coupling, i.e. for 
those that contain the momentum transfer $\vec Q$. A similar modification 
should also be made in the expression for $\beta^-$~(eq.~(\ref{beta})), 
but as that contribution is already very small compared to those from the 
spin-independent $\alpha$ terms, see Table~\ref{reltab}, it will not be 
considered in this paper.

The hyperfine interaction that is modeled as screened one-gluon exchange 
(OGE) between the quark and the antiquark also contains a central 
component which gives a contribution to the isovector exchange current. 
This contribution may be obtained as~\cite{TL2}:

\begin{equation}
\vec V_g=-{V_g(k)\over 2 m_q^2}\left[
{m_q\over M_Q}(\vec p_Q\,'+\vec p_Q+i\vec\sigma_Q\times\vec k\,)
+i\vec \sigma_q\times\vec k\right].
\label{gluecur}
\end{equation}
In the above expression for the OGE exchange current, the $M_Q^{-1}$ 
dependent term arises from 
the spatial current coupling
$\vec\gamma_q\cdot\vec\gamma_{\bar Q}$ term, and the 
second term from the 
charge coupling 
$\gamma^4_q\gamma^4_{\bar Q}$ term in the  
the OGE interaction. $V_g(k)$ denotes the form of the 
OGE interaction in momentum space, which is conveniently expressed as

\begin{equation}
V_g(k)=-{16\pi\over 3}{\alpha_{\mathrm S}(k^2)\over k^2},
\end{equation}
where a color factor of 4/3 has been included. In the interaction model 
used in ref.~\cite{TL2}, the running coupling constant was taken to have 
the following screened form:

\begin{equation}
\alpha_{\mathrm S}(k^2)={12\pi\over 27}{1\over \ln[(k^2+4 
m_g^2)/\Lambda_0^2]}, \label{alphas}
\end{equation}
where the parameters $m_g$ and $\Lambda_0$ were obtained as $m_g=240$ MeV 
and $\Lambda_0=280$ MeV respectively. In case of the $\pi\pi$ decay 
of the $D$ meson states with $L=1$, the term with the sum of charm quark 
momenta in eq.~(\ref{gluecur}), gives the largest contribution. As this 
term is inversely proportional to the mass of the charm quark, it is 
evident that the gluon exchange contribution will be smaller than the 
corresponding contribution from the exchange current~(\ref{ex3}) that is 
associated with the confining interaction. This main term of the isovector 
exchange current that is associated with 
the OGE interaction may be taken into account in the non-relativistic 
approximation by modifying eq.~(\ref{ex4}) to read

\begin{equation}
{1\over 2 f_\pi^2}\rightarrow{1\over 2 f_\pi^2}
\left(1-{V_c(r)\over m_q}+{V_g(r)\over M_Q}\right),
\label{exg}
\end{equation}
where $V_g(r)$ is the central component of the one-gluon exchange
potential. For a bare OGE interaction one would
have $V_g(r)=-4\alpha_{\mathrm S}/3r$. This simple expression is however 
significantly modified by the running coupling as given by 
eq.~(\ref{alphas}). This effect may be included as in ref.~\cite{TL2}, by 
expressing $V_g(r)$ as

\begin{equation}
V_g(r) = -\frac{4}{3}\frac{2}{\pi}\int_0^\infty dk\, j_0(kr) \,
\alpha_{\mathrm S}(k^2).
\end{equation}
This form reduces to the static OGE potential given above, if the running 
coupling $\alpha_{\mathrm S}$ is taken to be constant. Note that the 
relativistic corrections arising from the quark and antiquark spinors have 
not been included in the above expression, since it is more natural to 
place them in the expression for the matrix element instead. Moreover, 
comparison of these expressions with those obtained for the 
confining interaction shows that the net effect of the exchange
current that is associated with OGE will have the same sign as that, which 
is associated with the confining interaction. 

In the 
expressions~(\ref{ex3}) and~(\ref{gluecur}), one factor of $1/m_q$ 
represents the static 
limit of the propagator of the intermediate negative energy light 
constituent quark. In view of the results obtained in the previous 
section, this static limit is expected to give rise to a considerable 
overestimate of the two-quark amplitude contribution. A more realistic 
treatment may be obtained if the static propagator $1/m_q$ is replaced by 
the symmetrized form $4/(2m_q+E+E')$ as in ref.~\cite{TL3}. Thus in the 
relativistic case, the appropriate replacement of the Weinberg-Tomozawa 
term would be the following extension of the replacement~(\ref{exg}):

\begin{eqnarray}
{1\over 2 f_\pi^2}\!\!\!\! &\rightarrow& \!\!\!\!
{1\over 2 f_\pi^2}\left[
1-\sqrt{\frac{(E_c+M_Q)(E_c'+M_Q)}{4E_cE_c'}}\left\{
V_c\left(\frac{|\vec r + \vec r\,'|}{2}\right) \frac{4}{2m_q+E+E'}
\left(\!1-{P^2-q^2/4\over (E_c'+M_Q)(E_c+M_Q)}\!\right) \right.\right. 
\nonumber \\
&&\left.\left.-V_g \left(\frac{|\vec r + \vec r\,'|}{2}\right)
\frac{4m_q}{2m_q+E+E'}
\left(\frac{1}{E_c'+M_Q} + \frac{1}{E_c+M_Q}\right) \right\}\right]
\label{confrgl}
\end{eqnarray}
where the factors containing $E_c'$ and $E_c$ arise from the spinors of 
the heavy antiquark line in Fig.~\ref{Zdiagram}. To take the relativistic 
effects into account demands employment of the relativistic version of the
matrix element ${\cal M}_0^{\mathrm{rel}}$ also in the case of the 
exchange current contribution. First of all, the contribution to the decay 
rate from the $\alpha^-$ term, eq.~(\ref{ny2}), should be modified to take 
into account the two-quark contributions to the Weinberg-Tomozawa 
interaction. This can be accomplished by replacing eq.~(\ref{ny2}) with

\begin{equation}
|T|_{\alpha^-}^2 = 6\left[B^-\,\left(\frac{\omega_b - \omega_a}{2}
{\cal M}_{1-}^{\mathrm{rel}} - \frac{Q_fz}{2}
{\cal M}_0^{\mathrm{rel}}\right) + \frac{1}{2f_\pi^2}\frac{Q_fz}{2}
\left({\cal M}_0^{\mathrm{Conf}} - {\cal 
M}_0^{\mathrm{OGE}}\right)\right]^2.
\label{TB1}
\end{equation}
The appropriate forms for the relativistic two-quark matrix elements in 
the above equation can be obtained by multiplying the expression for 
${\cal M}_0^{\mathrm{rel}}$ with the relativistic factors included in 
eq.~(\ref{confrgl}) above. In addition, the argument $\vert \vec r + \vec 
r\,'\vert /2$ of the potentials will be approximated by the expression 
$\sqrt{(r^{'2} + r^2)/2}$. The resulting forms for the two-quark matrix 
elements in eq.~(\ref{TB1}) are then obtained as

\begin{eqnarray}
{\cal M}_0^{\mathrm{Conf}}\!\!\!&=&\!\!\!{1\over
3\pi}\int_{0}^{\infty}dr'r'\int_{0}^{\infty}dr\,r
\left[u'_0(r')u_1(r)-u_0(r')u'_1(r)-2{u_0(r')u_1(r)\over r}\right]
\int_{0}^{\infty} dP\,P^2 \int_{-1}^{1}dv\,f_{\mathrm{BS}}(P,v) 
\nonumber \\
&&V_c\left(\sqrt{\frac{r^{'2} + r^2}{2}}\,\right)\frac{4}{2m_q+E+E'}
\,\sqrt{\frac{(E+m_q)(E'+m_q)}{4EE'}}
\left(\frac{1}{E+m_q}+\frac{1}{E'+m_q}\right) 
\nonumber \\
&&\sqrt{\frac{(E_c+M_Q)(E_c'+M_Q)}{4E_cE_c'}}
\left(\!1-{P^2-q^2/4\over (E_c'+M_Q)(E_c+M_Q)}\!\right)
\:j_0\left(r'\sqrt{P^2+{q^2\over 16}-{Pqv\over 2}}\:\right)
\nonumber \\
&&j_0\left(r\sqrt{P^2+{q^2\over 16}+{Pqv\over 2}}\:\right),
\label{MCrel}
\end{eqnarray}
for the confining interaction, and

\begin{eqnarray}
{\cal M}_0^{\mathrm{OGE}}\!\!\!&=&\!\!\!{1\over
3\pi}\int_{0}^{\infty}dr'r'\int_{0}^{\infty}dr\,r
\left[u'_0(r')u_1(r)-u_0(r')u'_1(r)-2{u_0(r')u_1(r)\over r}\right]
\int_{0}^{\infty} dP\,P^2 \int_{-1}^{1}dv\,f_{\mathrm{BS}}(P,v) 
\nonumber \\
&&V_g\left(\sqrt{\frac{r^{'2} + r^2}{2}}\,\right)\frac{4m_q}{2m_q+E+E'}
\,\sqrt{\frac{(E+m_q)(E'+m_q)}{4EE'}}
\left(\frac{1}{E+m_q}+\frac{1}{E'+m_q}\right) 
\nonumber \\
&&\sqrt{\frac{(E_c+M_Q)(E_c'+M_Q)}{4E_cE_c'}}
\left(\frac{1}{E_c+M_Q}+\frac{1}{E_c'+M_Q}\right) 
\:j_0\left(r'\sqrt{P^2+{q^2\over 16}-{Pqv\over 2}}\:\right)
\nonumber \\
&&j_0\left(r\sqrt{P^2+{q^2\over 16}+{Pqv\over 2}}\:\right),
\label{Mgrel}
\end{eqnarray}
for the OGE interaction. The numerical results that follow when 
eq.~(\ref{TB1}) is employed are displayed for the most important decay 
modes in Table~\ref{tbtab}.

\begin{table}[h!]
\begin{center}
\begin{tabular}{l|c|c|c||c|c}
& \multicolumn{3}{c||}{$|T|_{\alpha^-}^2$} & \multicolumn{2}{c}{Total} \\
\quad Decay & Rel & +Conf & +OGE & $g_A^q = 1$ & $g_A^q = 0.87$ \\ 
\hline\hline &&&&& \\
$D_2^*\rightarrow D^*\pi\pi$ & 2.864 & 2.377 & 2.144 & 3.05 MeV & 1.82 MeV \\
$D_1\rightarrow D^*\pi\pi$   & 1.291 & 1.076 & 0.974 & 1.34 MeV & 0.80 MeV \\ 
$D_1^*\rightarrow D\pi\pi$   & 7.915 & 6.535 & 5.872 & 8.62 MeV & 5.17 MeV \\
&&&&& \\
\end{tabular}
\caption{Numerical results for the most important $\pi\pi$ decay modes 
obtained upon employment of the two-quark contributions to the 
Weinberg-Tomozawa interaction. The modifications to the $|T|_{\alpha^-}^2$ 
amplitude with $g_A^q = 1$ are shown as follows: The column "Rel" gives 
the one-quark result for each decay rate (cf. Table~\ref{reltab}), in the 
column "+Conf\," 
the contribution from ${\cal M}_0^{\mathrm{Conf}}$ has been added, and in 
"+OGE", the results that follow when both ${\cal M}_0^{\mathrm{Conf}}$ and 
${\cal M}_0^{\mathrm{OGE}}$ are employed are given. The resulting total 
decay widths for $\pi\pi$ decay are also shown, for both $g_A^q = 1$ and  
$g_A^q = 0.87$.}
\label{tbtab}
\end{center}
\end{table}

\section{Discussion}

In the absence at the present time of experimental data 
on the two-pion decay widths  of the $D$ mesons, it
is instructive to compare the calculated results
with the empirical knowledge of the analogous
pionic decay modes of the positive parity strange mesons.
The $K_2^*(1430)$ state, which is the 
strange analog of the $D_2^*(2460)$ has a width of
$\sim$ 13 MeV for $\pi\pi$ decay~\cite{PDG}. Given the smaller
phase space and much narrower radial wavefunctions, the present calculated 
$\pi\pi$ decay width value of 3.1 MeV for the $D_2^*(2460)$ 
hence appears to be reasonable. Furthermore, in 
ref.~\cite{PDG}, the branching ratio $K_2^* \rightarrow K^*\pi$ is given 
as $(24.7 \pm 1.5)\%$ while that for $K_2^* \rightarrow K^*\pi\pi$ is 
reported to be $(13.4 \pm 2.2)\%$. Thus, the available experimental data 
indicates that for this particular decay mode, the width for $\pi\pi$ 
decay ought to be $\sim 55 \%$ of the width for $\pi$ decay. If the 
current calculation is compared to that in ref.~\cite{TL3}, the ratio of 
$D_2^* \rightarrow D^*\pi\pi$ to $D_2^* \rightarrow D^*\pi$ is obtained 
as $\sim 60 \%$ for $g_A^q = 1$. One may thus conclude that a width for 
$D_2^* \rightarrow D^*\pi\pi$ of 3 MeV, as obtained in the current 
calculation, is what would be expected by comparison with the 
corresponding decays of the strange $K$ mesons. The present results also 
appear realistic in view of the fact that the empirical nonobservation
of the decay mode $K_2^*\rightarrow K\pi\pi$ indicates 
that the amplitude for two-pion decay is mainly of spin-independent 
character. 

The strange analog 
of the $D_1(2420)$ state is most likely a mixture of the $K_1(1270)$ and 
the $K_1(1400)$ states. The two-pion decay modes of these two states are 
dominated by $\rho$ meson decay. As $\rho-$meson decay is kinematically
impossible for the $D_1(2420)$ as well as the other $L=1$ $D$ mesons, 
these decay modes have little bearing on the decays of the latter. 
Addition of the calculated decay widths for single- and two-pion decay 
gives predictions for the total widths 
of the $L=1$ $D$ meson states. These results are shown in Table~\ref{tottab}.

\begin{table}[h!]
\begin{center}
\begin{tabular}{c|r|r|c|c|c}
$D$ meson state & $\pi$ width & $\pi\pi$ width & Total $g_A^q=1$ & Total 
$g_A^q=0.87$ & Experiment \\ \hline\hline &&&&& \\
$D_2^*$ & 15.7 MeV & 3.05 MeV       & 18.8 MeV & 13.7 MeV & 
25$^{+8}_{-7}$ MeV \\ &&&&& \\
$D_1$   & 13.6 MeV & 1.34 MeV       & 14.9 MeV & 11.1 MeV & 
$18.9_{-3.5}^{+4.6}$ MeV \\ &&&&& \\
$D_0^*$ & 27.7 MeV & $\sim$ 0.1 MeV & 28.8 MeV & 21.0 MeV & -- \\ &&&& \\
$D_1^*$ & 13.2 MeV & 8.62 MeV       & 21.8 MeV & 15.2 MeV & -- \\ &&&& \\
\end{tabular}
\caption{Total decay widths for strong decay of the $L=1$ $D$ mesons that 
follow when the results of the present calculation of the $\pi\pi$ decay 
widths are added to those for single pion decay obtained in 
ref.~\cite{TL3}. The individual results for $\pi$ and $\pi\pi$ decay are 
shown for $g_A^q = 1$. The experimental results, when available, have 
been taken from ref.~\cite{PDG}. Note that the $D_1$ is a spin triplet 
state with $J=1$ and $D_1^*$ is the corresponding spin singlet state.} 
\label{tottab} 
\end{center} 
\end{table}

As indicated by Table~\ref{tottab}, the present calculation of the two 
pion decay widths of the $L=1$ charm mesons completes the calculation of 
the corresponding single pion decay widths of ref.~\cite{TL3}. The 
calculations show that, if the charm mesons are described as two-particle 
systems formed of one light flavor constituent quark and one charm 
antiquark, where only the light quark couples to pions, a fair description
of the presently known decay widths of these mesons may be obtained. Given 
the chiral form for the coupling between the pions
and the light constituent quarks, a better description
of the total pionic decay widths obtains with the value $g_A^q = 1$ than 
with smaller values, even though the empirical data are still very crude 
at this time. If other decay modes, such as $\pi\pi\pi$ and $\eta$ 
decay should turn out to give appreciable contributions to 
the total decay widths of the positive parity charm mesons, a somewhat 
smaller value for $g_A^q$ may be favored. However, as $\pi\pi\pi$ decay 
has so far not been detected for the strange $K_2^*$ meson~\cite{PDG} and 
since the $\eta$ decay width of that same state is empirically found 
to be very small, that possibility is apparently not supported by 
experiment at this time. From the calculations of ref.~\cite{Godfrey}, one 
may conclude that the width of the decay mode $D_2^*\rightarrow D\eta$ is 
probably less than 0.25 MeV.

Furthermore there is the possibility of two-pion decay of the $D_2^*$ and 
$D_1$ mesons through an intermediate $D_1^*$ which is close to its mass 
shell, an effect which has been investigated by ref.~\cite{Falk}. There, 
this mechanism was found to contribute significantly to the two-pion decay 
widths, although the effect is very sensitive to the widths and spin-orbit 
splittings of the $L=1$ $D$ mesons and is thus very difficult to estimate.

Finally it should be noted that the results are quite sensitive to the 
exact form of the $B^-$ amplitude, as given by eq.~(\ref{Bamp-}). 
Therefore, crude approximations have to be avoided in the case of this 
amplitude. If the $B^-$ amplitude is treated non-relativistically, which 
implies dropping the nonlocal and $\vec Q$ dependent terms altogether, 
large overestimates of that contribution to the $\pi\pi$ decay widths will 
result.

\vspace{1cm}
\centerline{\bf Acknowledgments}
\vspace{1cm}
We are indebted to Professor M. Robilotta for 
instructive discussions. DOR thanks the W. K. Kellogg
Radiation Laboratory of the California Institute
of Technology for its hospitality during the
completion of this work.
Research supported in part by the Academy of Finland
by grant No. 43982. 


\end{document}